\documentclass[a4paper,11pt]{article}
\usepackage{jcappub} % for details on the use of the package, please
\usepackage[T1]{fontenc} % if needed

\title{\boldmath Direct detection of the cosmic expansion: the redshift drift and the flux drift}

\author[a]{Krzysztof Bolejko,}
\author[b]{Chengyi Wang,}
\author[c]{Geraint F. Lewis}

\affiliation[a]{School of Natural Sciences, College of Sciences and Engineering, University of Tasmania, Private Bag 37, Hobart TAS 7001, Australia}
\affiliation[b]{School of Astronomy and Space Science, Nanjing University, Xianlin Avenue 163, Qixia District, Nanjing Shi, Jiangsu Sheng, 210046, China}
\affiliation[c]{Sydney Institute for Astronomy, School of Physics, A28, The University of Sydney, NSW, 2006, Australia}

\emailAdd{krzysztof.bolejko@utas.edu.au}

\abstract{The redshift drift, the change of cosmological redshift with time, is a direct consequence of the expansion of the Universe. Thus the measurement of the cosmological redshift drift will offer a direct test of our models of cosmology. The magnitude of the effect is very small, i.e. the spectral shift is of order of $10^{-10} - 10^{-9}$ over the period of a decade, but the next generation facilities such as ELT and SKA will be able to directly detect the expansion of our Universe by the year 2040. 
In this paper we focus on detectebility of this effect, including strategies of overcoming the kinematic contamination of the cosmological signal. We also show that the redshift drift directly impacts the change of flux.   Thus apart from the {\em redshift drift}, measurements of the {\em flux drift} will provide an additional tool of detecting the expansion of the universe, including its acceleration. We discuss the strategies of detecting the 
 {\em flux drift} and show that by including the  {\em flux drift} signal to the {\em redshift drift} signal we boost the chances of a direct detection of the expansion of the Universe. We show that if only the stability of flux is at the level of $\Delta F/F \sim 10^{-6}$ then the SKA1-mid Array should be able to detect these effects, before the ETL and the full SKA. Thus, by including the {\em flux drift} into the SKA1-mid Array's analysis pipeline, we could be able to provide by mid-2030s a direct evidence of the expansion of the universe including its accelerating phase.  }

\begin{document}
\maketitle
\flushbottom

\section{Introduction}

The last century has seen a revolution in our understanding of the cosmos, beginning with the development of the mathematical framework of relativistic cosmologies and the discovery of the expansion of the universe, to the realization that this expansion is accelerating due to the dominating presence of dark energy. 
One direct test of our cosmological models is the redshift drift, the change of a distant objects redshift over time, commonly referred to as the Sandage-Loeb effect \citep{1962ApJ...136..319S,1998ApJ...499L.111L}, although this idea has surfaced a number of times in the literature over the years~\citep{1962ApJ...136..334M}.
While the magnitude of the redshift drift is expected to be small recent studies have demonstrated that it should be in reach with at least a decades worth of observing with the next generation of telescope and instruments~\citep{2007MNRAS.382.1623B}.
However, a number of kinematic sources can contaminate the redshift drift signal \citep{2010MNRAS.402..650K}, and these must be fully accounted in an observational program to realise the redshift drift as a probe of cosmology.

The advent of the Square Kilometre Array (SKA) offers a new window through the accurate detection of a large number of high redshift sources. In this paper, we will consider the detectability of the redshift drift with the SKA, including the flux drift. We consider the theoretical background for the redshift drift in Section.~\ref{background}. In Section.~\ref{measure} we discuss strategies of measuring the redshift drift including ways of handling the kinematic contamination. In Section~\ref{fluxdrift} we discuss the flux drift and its detectibility. 
We present our conclusions in Section~\ref{conclusions}.

\section{Redshift drift}\label{background}

Due to cosmic expansion observable properties of 
cosmological objects change with time. 
The most studied effect is the redshift drift \citep{2007MNRAS.382.1623B,2008PhRvD..77b1301U}, but other properties include position drift \citep{2010PhRvD..81d3522Q,2011PhRvD..83h3503K} as well as the distance drift \citep{2018JCAP...03..012K,2018arXiv181110284G}.

\begin{figure}[h!]
\centering
\includegraphics[scale=1.0]{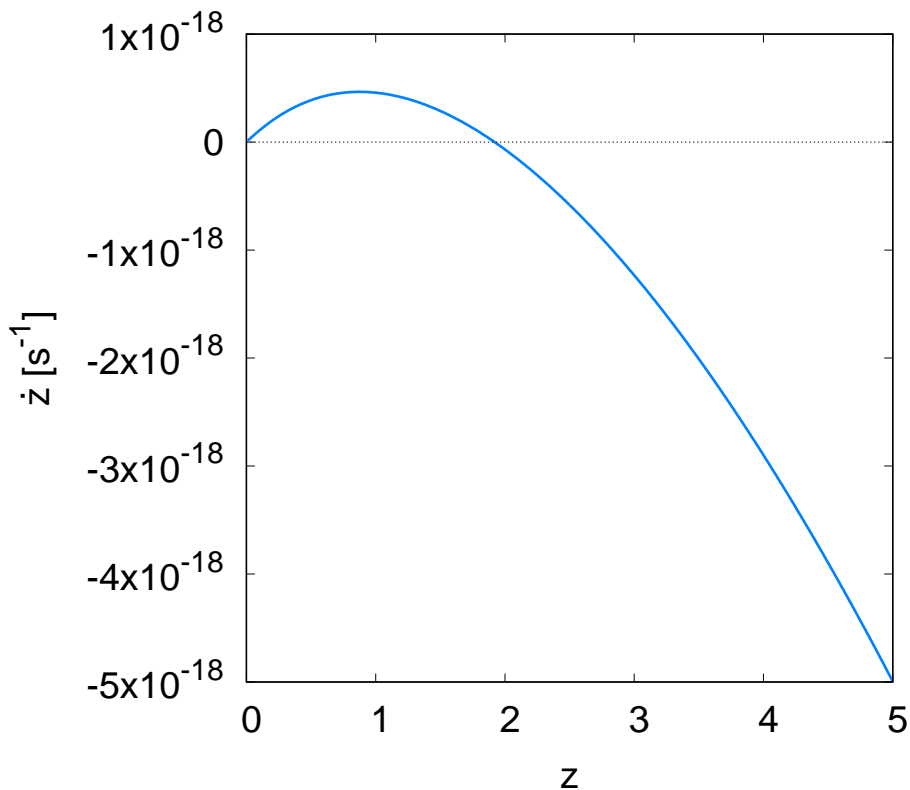}
\caption{\label{fig1}The cosmological redshift drift signal as a function of redshift (eq.~\ref{redshiftlcdm}). Note that the redshift drift is positive for low redshifts, passing through zero at $z\sim2$, and becoming increasingly negative at high redshift.}
\end{figure}

The redshift drift for FLRW models is given by a simple relationship
\begin{equation}
\dot{z} = H_0 (1+z) - H(z),
\end{equation}
where $z$ is the redshift of the source, $H_0$ is the present day value of the Hubble constant, and $H(z)$ is the Hubble constant at the redshift of the source. For a flat universe with a present day matter density of $\Omega_m$ and cosmological constant $\Omega_\Lambda$, this can be written as
\begin{equation}
    \dot{z} = H_0 \left ( (1+z) - \sqrt{ \Omega_m (1+z)^3 + \Omega_\Lambda } \right).
\label{redshiftlcdm}
\end{equation}
In Figure~\ref{fig1}, we present $\dot{z}$ of a fiducial universe, with $\Omega_m=0.3$ and $\Omega_\Lambda=0.7$, demonstrating that this is positive at redshifts less than $z\sim2$, and then negative at high redshifts; this relationship demonstrates that our universe can be thought of as transitioning from a decelerating phase in the past, when it was dominated by matter, into an accelerating phase due to the presence of dark energy. However, this also reveals that the redshift drift signal in the nearby universe is largest at $z\sim1.2$, but grows substantially more negative in the high redshift universe. Hence, the redshift drift will be most prominent for the more distant sources in the universe~\citep{2015APh....62..195K}. 

\section{Measuring the redshift drift} \label{measure}

\subsection{Single-object optical spectrocopy}

The simplest strategy to measure the drift is  to determine a source redshift at one instant and then repeat the measurement after $\Delta t$ 
\begin{equation}
\Delta z = z (t_2) - z (t_1) = \dot{z} \, \Delta t+ \Delta z_{sys},
\label{obsreddrift}
\end{equation}
where $\dot{z}$ is the redshift drift and $\Delta z_{sys}$ are sources of systematic velocity.
For example, after 10 years the expected change in redshift due to cosmological expansion, at $z \sim 1 $ is
\[\dot{z} \, \Delta t = \Delta z = 10^{-10}.\]
Thus, in order to measure such a change in a single experiment the accuracy of at least $\Delta z \sim 10^{-10}$ is required, and the
proposed CODEX experiment will achieve  such a precision \cite{CODEX}. However, this will also require 
keeping all the sources of systematic velocities below the level $\Delta z_{sys} \sim 10^{-10}$, which includes the stability of the instrument and modelling the motion of the observatory up to a few cm/s.

\subsubsection{Kinematic Contamination}\label{doppler}

\begin{figure}
\centering
\includegraphics[scale=0.65]{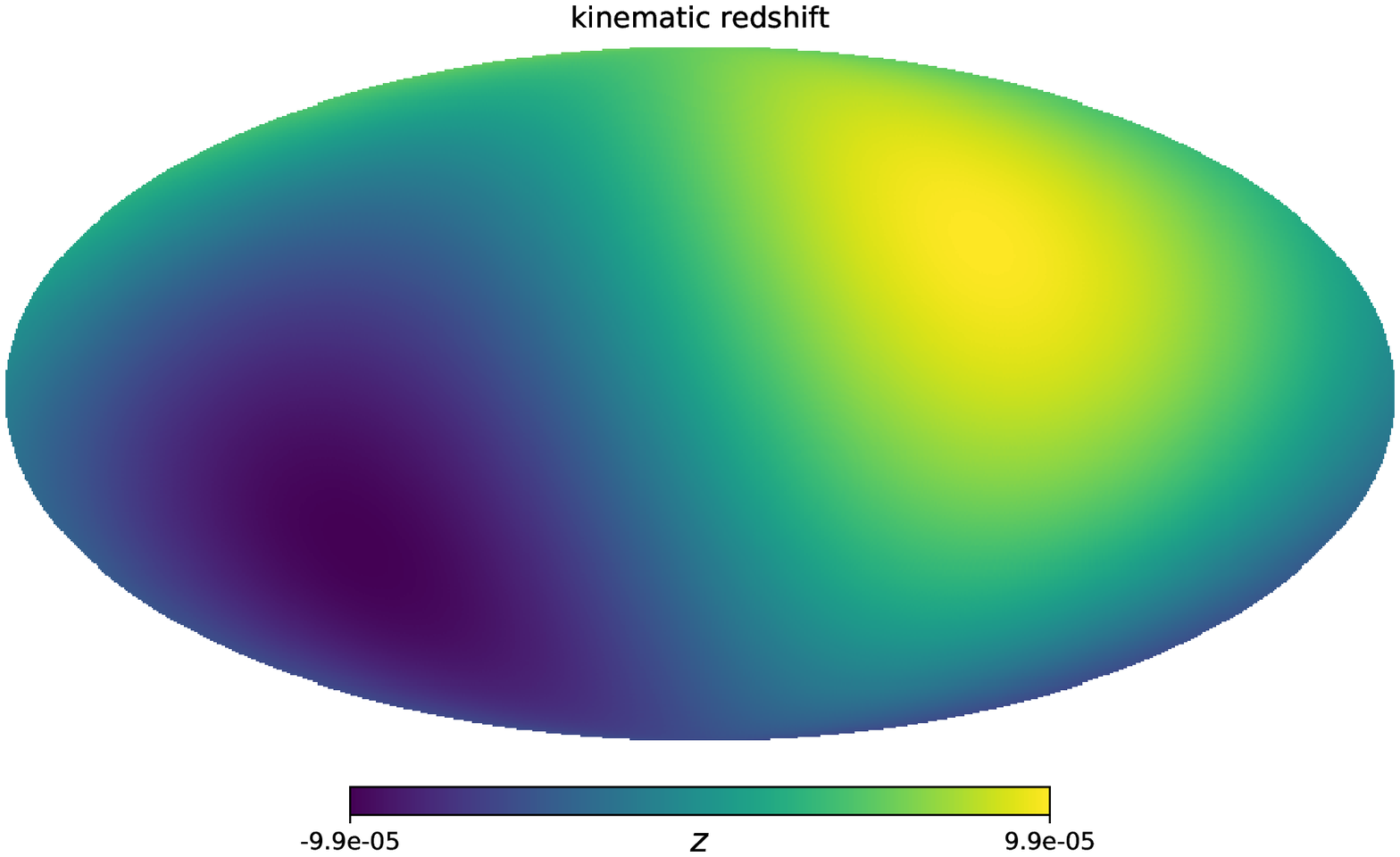}
\includegraphics[scale=0.65]{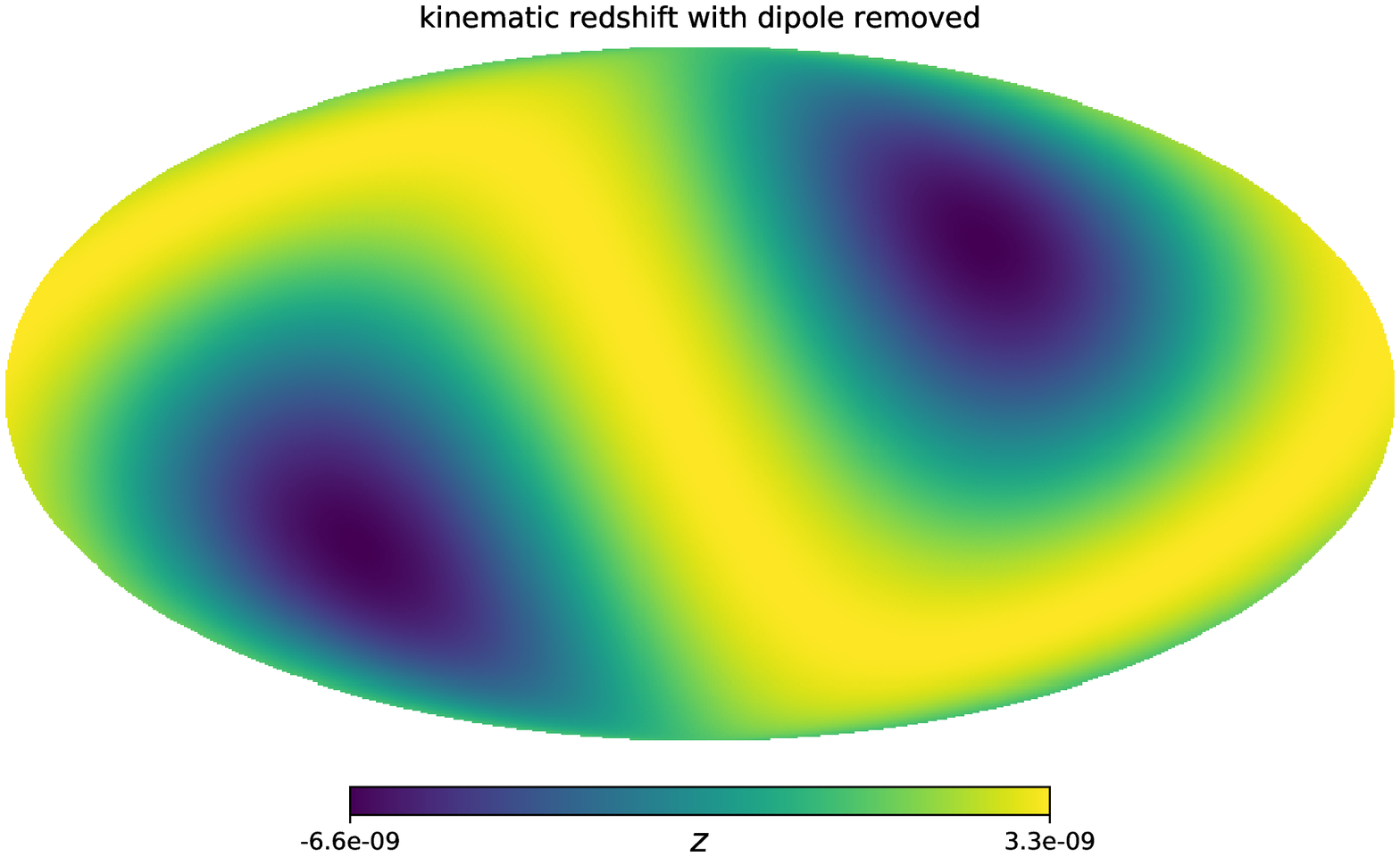}
\caption{\label{fig:2}The kinematic redshift induced by the Earth's orbital motion as a function of angular position (upper panel) and with the dipole removed (lower panel).}
\end{figure}

The motion of the observatory contributes to redshift via the Doppler effect

\begin{equation}
1 + z = \gamma (1+ v_i \,e^i),
\end{equation}
where $v^i$ is the {\em i-th} component of the velocity vector (noting that speed of light is $c=1$), $e^i$ is the {\em i-th} component of the vector pointing in the direction to the source; $v_i e^i \equiv  \sum\limits_{i=1}^{i=3} e^i v^i$;  $\gamma^{-2} = 1 - \beta^2$ and $\beta^2 = v_i v^i \equiv  \sum\limits_{i=1}^{i=3} v^i v^i$.
In addition, the motion affects the direction via the aberration \cite{2002PhRvD..65j3001C}

\begin{equation}
e^i =  \left( \frac{n^j v_j + \beta}{1+ n^j v_j} \right) v^i +
\frac{ n^i - n^j v_j v^i}{\gamma (1 + n^j v_j)},
\end{equation}
where $n^i$ is the direction for a stationary observer. In the 1st order in $\beta$, this reduces to

\begin{equation}
e^i =  n^i \, ( 1 - v_j \, n^j ),
\end{equation}
and the  redshift is 
\begin{equation}
1 + z = \gamma (1+ v_i n^i - (v_i n^i )^2),
\end{equation}
where $v_i n^i$ produces the dipole and $ (v_i n^i )^2$ gives rise to a quadruple.
Figure \ref{fig:2} shows the kinematic contamination of the redshift induced by 
 Earth's orbital motion. The significance of these results are that even if we model accurately Earth's orbital motion but only account for the dipole then we still have a 
 kinematic contamination due to the aberration effect, leading to $\Delta z_{sys} \sim 10^{-9}$. Several additional sources of kinematic contamination are present, considering potential motion of an observer, and are presented
 in Table~\ref{Table}. These poses a challenge as there are several effects that can induce  the Doppler shift comparable with the expected signal from the cosmological redshift drift. For example: heliocentric redshifts are typically modelled with a precision of $10^{-4}$ giving residual velocities of order  $\Delta v \sim {\rm km}/{\rm s}$; correction to Earth's motion due to perturbations from other planets are also of order meters per second. In addition such effects as the aberration effect or tidal motion or plates tectonics need also be taken into account.

If one can model all these effects with a precision of a few cm/s that would translate to uncertainty of $\Delta z = 10^{-10}$. To increase the accuracy further one needs to increase the number of objects to average out at least some of systematics.
With a 40-m class telescope, available with the planned advent of the ELT, one would require a 4000 hours of observing time to get at least 10 objects with spectra measured with accuracy of $\Delta z = 10^{-10}$ \citep{2008Msngr.133...10L}. 
However, if the uncertainty of measuring the motion of the spectrograph cannot reach the precision of few cm/s but is of order of 100 m/s then assuming that the uncertainties are uncorrelated
\[ S/N \sim \frac{10^{-10}}{ 10^{-7}} \sqrt{N_{QSO}}, \]
one would require at least $10^6$ quasar spectra which is not achievable as it would take decades of dedicated time of a 40m-class telescope solely for the purpose of such an experiment.

\begin{table}
\caption{Contribution to the redshift drift from various effects.}
\label{Table}
\begin{center}
\begin{tabular}{ l|c } 
  { } & $\Delta z_{sys}$  \\ 
  \hline
  %%%%%%%%%%%%%%%%%%%%%%%%%%%%%%%%%%%%%%%
Earth's orbital motion (dipole) & $10^{-4}$  \\
Earth's rotation at latitude $32^\circ$ & $10^{-6}$  \\
Earth's orbital motion (quadruple) & $10^{-9}$  \\
Tidal motion & $10^{-10}$  \\
Plate tectonics & $10^{-10}$  \\
Sun's motion within the galaxy & $10^{-10}$  \\
Galaxy's motion  &  $10^{-13}$  \\
 \hline
\end{tabular}
\end{center}
\end{table}

\begin{figure}[h!]
\centering
\includegraphics[scale=1.0]{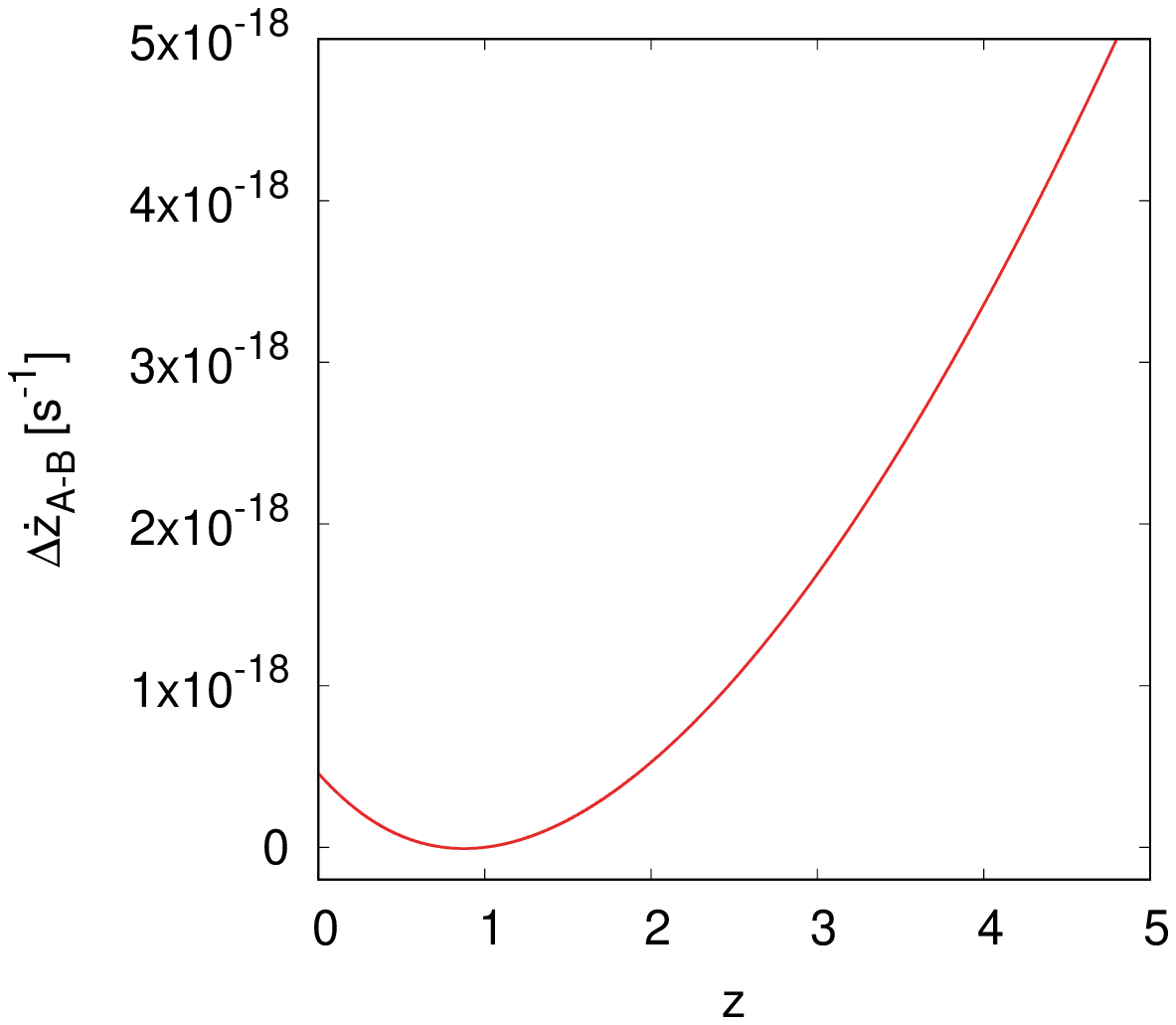}
\caption{\label{fig:3}The difference between the redshift drift at redshift $z_A$ and $z_B$, i.e. $\Delta \dot{z}_{A-B} = \dot{z}_A- \dot{z}_B$, where $z_A = 1$ and $z_B = z$. }
\end{figure}

\subsection{Multi-object optical spectroscopy}

If modelling the spectograph's motion with the accuracy of cm/s cannot be acheived, then a more efficient strategy is to target simultaneously a pair of objects at two different redshift. Measuring simultaneously redshift of object $A$ and $B$ and taking a difference will remove the kinematic contamination (as long as the angular separation is small)

\begin{equation}
\Delta z_{A-B} (t_2) - \Delta z_{A-B} (t_1) = \Delta \dot{z}_{A-B} \, \Delta t.
\end{equation}

\begin{figure}
\centering
\includegraphics[scale=1.3]{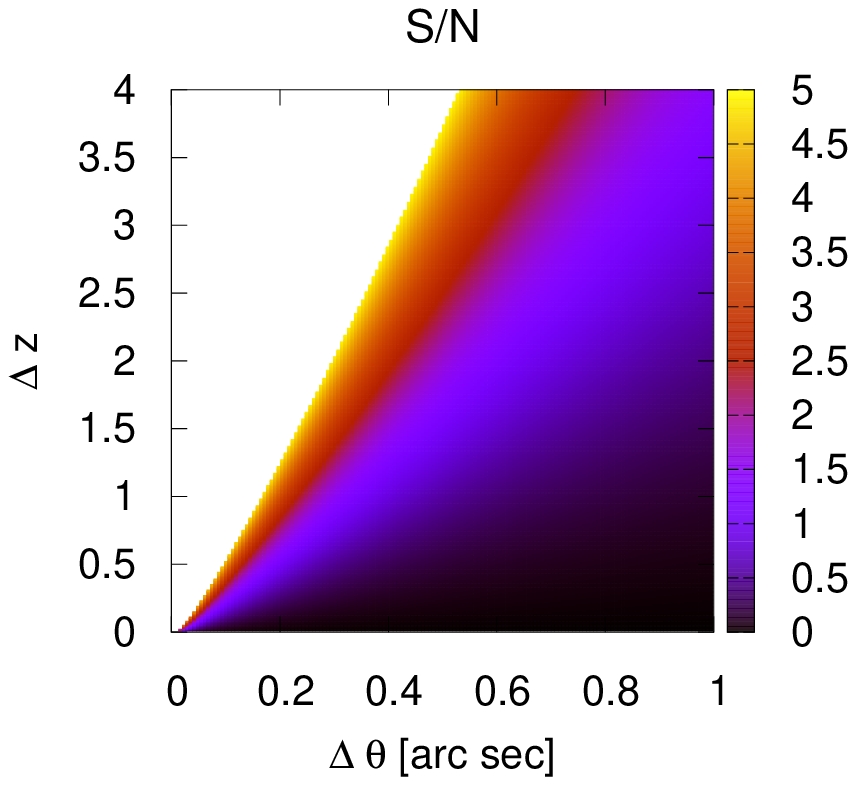}
\caption{\label{fig:snzt} Signal to noise for a measurement of a $\Delta \dot{z}_{A-B}$ as a function of angular and redshift separation for a CODEX type experiment on 40m-class optical telescopes.}
\end{figure}

Assuming that $z_A = 1$, the expected signal for such a measurement is presented in Fig.~\ref{fig:3}. The presented signal assumes a perfect alignment. If the sources are separated by an angle $\Delta \theta$ then the Doppler contamination will no longer be 
the same for both objects $A$ and $B$ (cf. Fig.~\ref{fig:2}). Figure \ref{fig:snzt} shows how the signal changes with the angular separation between the observed objects. The noise is modelled as the Doppler contamination of an amplitude of 100 m/s. 

As seen from Fig. \ref{fig:snzt}, with the redshift separation approximately 1, $\Delta z = z_A - z_B \approx 1$, and the angular separation is less than $0.2$ arc sec one could detect the redshift drift without the need for ultra-accurate modelling of the motion of the observatory. With 40-m size telescope such a project could be completed with a few thousand hours. After 10 years, the survey of the same objects would need to be repeated resulting in detection of the redshift drift.

\subsection{Measuring the redshift drift with SKA} \label{measureSKA}

The frequency resolution of SKA  is 1 KHz at 1 GHz \citep{SKA}. We will assume that redshift will be measured with similar precision, hence $\Delta z = 10^{-6}$.
With a few antennas pointing across the sky one could measure the CMB temperature  \cite{2000MNRAS.315..808S}
and infer the kinematic dipole at the level of $10^{-5} - 10^{-6}$ precision directly bypassing the  need for theoretical modelling of Earth's orbital motion (cf. Fig. \ref{fig:2}). With the dipole measure accurately with the precision of $10^{-5} - 10^{-6}$ the aberration effects can also be accurately inferred. Additional kinematic contamination due to Earth's rotation, tides, and plate tectonics can also be accounted for with the precision of cm/s and thus the observed signal can be corrected for kinematic contamination as the measurements are taken.  Still to measure the redshift drift with the SKA that has a precision of $\Delta z = 10^{-6}$ will require at least $10^7$ per redshift sources for the uncertainties to average out \cite{2015aska.confE..27K}. 

The number of HI galaxies observed by the SKA will depend on the flux sensitivity. 
Based on the numerical simulations the 
expected number of HI galaxies per sq. degree can be modelled as \cite{2009ApJ...703.1890O}
\begin{equation}
    \frac{ {\rm d} N} {{\rm d} z} = 10^{c_1} \, z^{c_2} \, e^{-c_3 z},
    \label{HIska}
\end{equation}
where $(c_1,c_2,c_3) = (5.75, 1.14,3.95)$ for the limiting peak flux density of $10$ $\mu$J and $(c_1,c_2,c_3) = (4.56, 0.43,6.86)$ for the limiting peak flux density of $100$ $\mu$J \cite{2009ApJ...703.1890O}. 
The flux sensitivity can be modelled as \cite{2015aska.confE..27K}

\begin{equation}
    \Delta I = \frac{ {\rm SEFD} } { \eta_s \sqrt{t \Delta \nu}},
\end{equation}
where $\eta_s = 0.9$, and the system equivalent flux density is
\begin{equation}
    {\rm SEFD} = \frac{2 k_{\rm B} T_{{\rm sys}}}{A_{{\rm eff}}},
\end{equation}
where $k_{\rm B}$ is the Boltzmann constant, $T_{{\rm sys}}$ is system temperature, and $A_{{\rm eff}}$ is the effective collecting area. 

\begin{figure}
\centering
\includegraphics[scale=0.9]{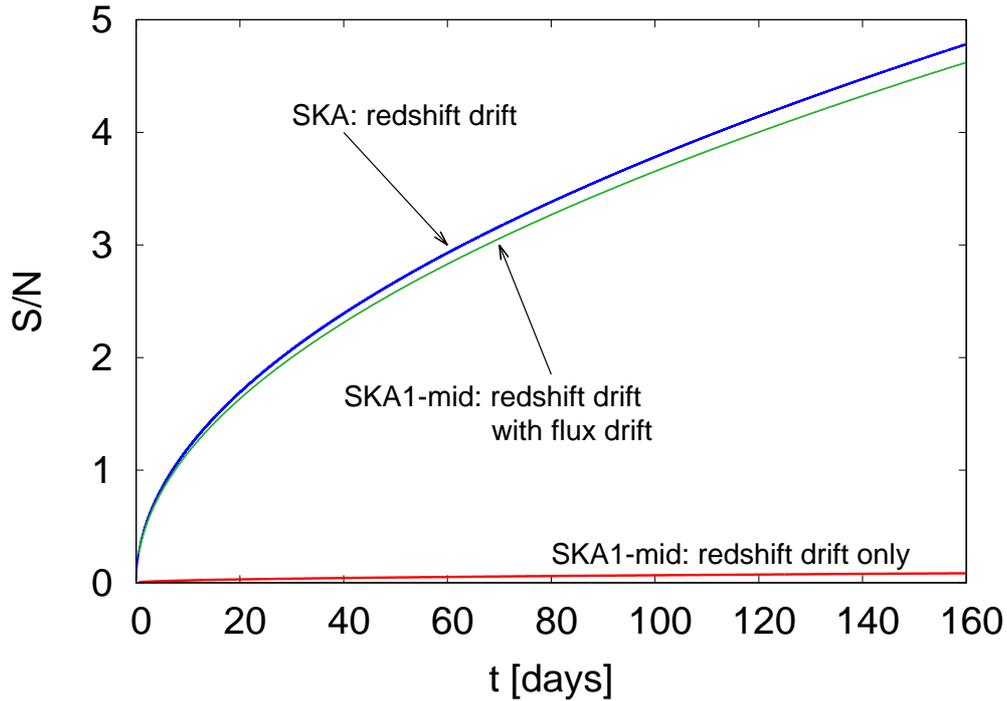}
\caption{\label{fig5} Time required to complete {\em Phase 1}. In order to measure the cosmological redshift drift, 10 after the phase 1 is completed the same {\em Phase 2} needs to be repeated. Thus, the total length of the SKA project to detect the cosmological redshift drift is $t = 10 {\rm ~years} + 2\times t_{Phase~1}$. With the SKA1-mid Array phase 1 would take approximately 40 years leading to a century long project to detect the cosmological redshift drift. With full SKA the whole project could be completed within the period of approximately 10 years. If the flux drift is included in the analysis of the data then the the SKA1-mid Array is capable of direct detection of cosmological expansion (assuming the stability of radio flux at the level of $10^{-6}$).}
\end{figure}

As in the case of the optical survey on ELT, the strategy of detecting the cosmological redshift drift with SKA is also based on 2 observing {\em Phases}. In each of the observing phases the spectra of HI galaxies are measured. {\em Phase 2} follows 10 yeas after {\em Phase 1} is completed and then the redshift drift is obtained from eq.~(\ref{obsreddrift}).
Figure~\ref{fig5} shows how long 1 observing phase would last with the full SKA and SKA1-mid Array.
For the full SKA with $A_{{\rm eff}}/T_{{\rm sys}} = 13000$ m$^2$ K$^{-1}$ and
2hr integration time and the 20 sq. degrees filed of view per pointing, 
 {\em Phase 1} could be completed within a year of operation \cite{2015aska.confE..27K}.
For the SKA1-mid Array with 
$A_{{\rm eff}}/T_{{\rm sys}} = 1300$ m$^2$ K$^{-1}$ and
12hr integration time and the 1 sq. degrees filed of view per pointing, {\em Phase 1}
would require more than 40 years of observing time to complete \cite{2015aska.confE..27K}.

\section{Redshift drift and flux drift}\label{fluxdrift}

\begin{figure}[h!]
\centering
\includegraphics[scale=1.0]{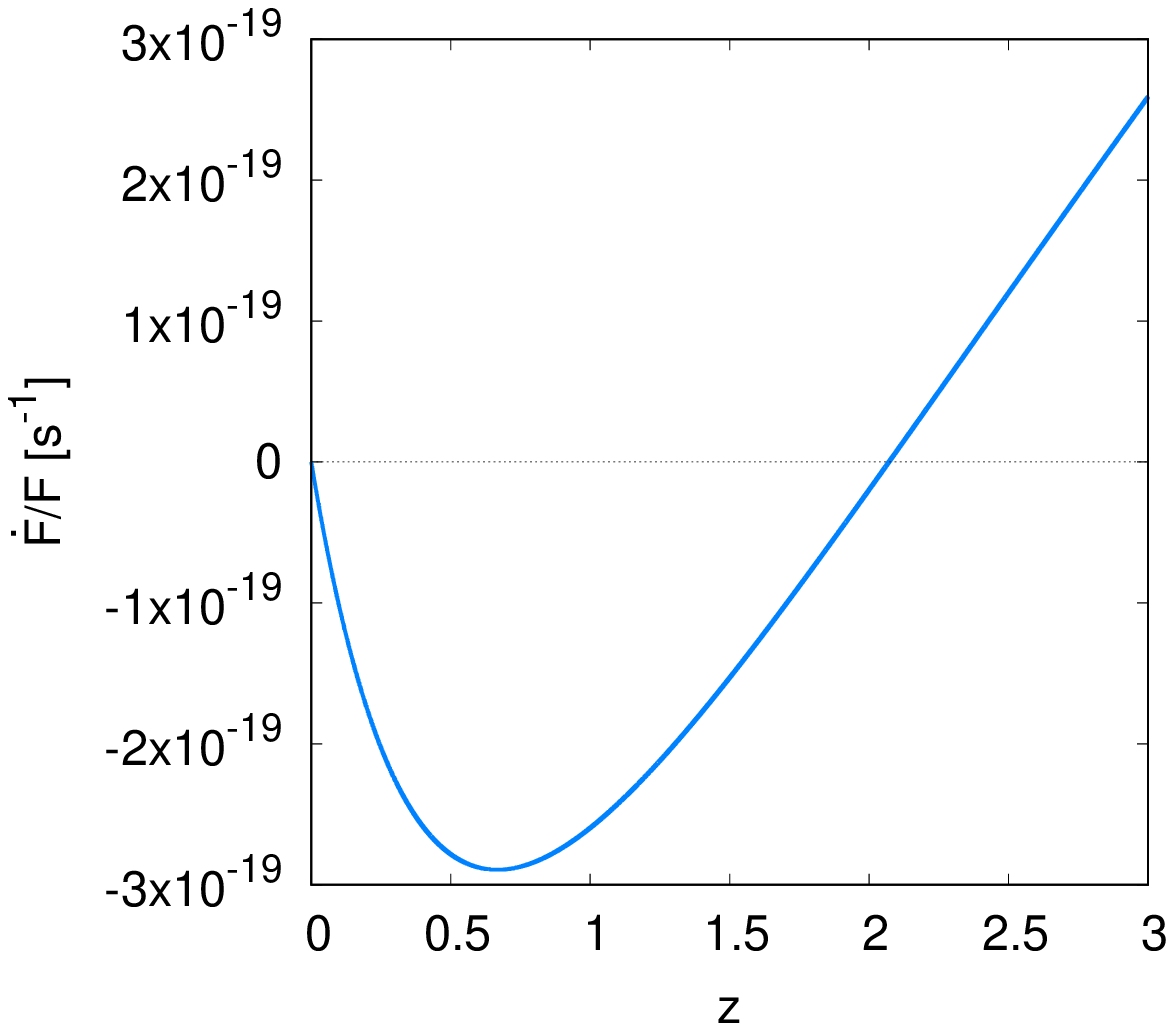}
\caption{\label{fig6}The cosmological flux drift signal as a function of redshift.  Note that the flux drift changes the sign at $z\sim2$ just as the redshift drift, but it peaks at a slightly lower redshift.}
\end{figure}

Apart from the redshift drift, radio observations also provide the opportunity to detect the {\em flux drift}.
The first person to investing the flux drift was Sandage \cite{1962ApJ...136..319S} who apart from the redshift drift also considered the bolometric flux drift. Below we focus on the specific flux, i.e. flux at a particular frequency. For radio observations, due to the steepness of the flux-frequency relation, the shift of the specific flux is more relevant then the change of the bolometric flux. Similarity the flux drift of the radio part of the spectrum is more significant than of the optical part, where the spectrum is less steep. This follows from the fact that the spectrum of radio sources can be modelled as a power law $\nu^{-\alpha}$ and so a shift in frequency is directly correlated with the shift of flux. The observed $\nu_o$ frequency is related to the emitted frequency $\nu_e$ via the redshift relation

\begin{equation}
   \nu_o = \frac{\nu_e}{1+z},
\end{equation}
and thus the frequency change due to the redshift drift  is 
\begin{equation}
   \Delta \nu = \nu_o(t_2) - \nu_o(t_1) =  - \nu_o \frac{\Delta z}{1+z}.
   \end{equation}
If ${\cal F}(\nu)$ is the source spectrum then the rate at which the source emits radiation 
in the frequency range $\nu$, $\nu + {\rm d} \nu$ is $L {\cal F}(\nu) {\rm d} \nu$.
The flux measured in the frequency range $\nu$, $\nu + {\rm d} \nu$ by the observer is \cite{Ellis2009}
\begin{equation}
F_\nu \, {\rm d} \nu_o = \frac{L}{4\pi d_L^2}  {\cal F} (\nu_e) {\rm d} \nu_e  =
\frac{L}{4\pi d_C^2} \frac{ {\cal F} (\nu_o (1+z)) }{(1+z)}  {\rm d} \nu_o,
\end{equation}
where $d_C$ is the comoving distance. 
For a power-law ${\cal F} \sim \nu^{-\alpha}$ the specific flux is
\begin{equation}
F_\nu = \frac{L}{4\pi d_C^2}   \frac{ \nu^{-\alpha}   }{ (1+z)^{1+\alpha} },
\end{equation}
and so the {\em flux drift} can be written as 
\begin{equation}
\frac{\Delta F_\nu}{F_\nu} = - \alpha  \frac{ \Delta z  }{ 1+z }.
\end{equation}
The flux drift as a function of redshift is presented in Fig.~\ref{fig6}. It provides a new opportunity to measure the cosmic expansion. With 
the flux drift one does not need to measure the spectral lines and infer the change of redshift, here one is only required to measure the flux at each channel across the bandwidth. 
However, measuring the flux drift requires observing galaxies that are intrinsically stable on the timescales of 10 years. While active galactic nuclei (AGN) are intrinsically variable and not suitable for measuring the flux drift, the radio galaxies, whose emission arises from large scale synchrotron emission, can be considered stable. Selecting target galaxies based on stability of their flux (in {\em Phase 1}) and excluding galaxies with any variability larger than  $10^{-6}$ (i.e. precision of the measurement) will provide a sample suitable to measure the flux drift along the side with the redshift drift. Any random variability below this level is expected to average out. 

The flux drift allows for the opportunity to directly detect the effect of the cosmic expansion before the completion of the full SKA, with 
the SKA1-mid Array. 
At frequencies 800 MHz the expected precision of measuring a signal from a radio galaxy for the SKA1-mid Array is \citep{SKA}
\[ \frac{\Delta F_\nu}{F_\nu} = 10^{-6}. \]
The SKA1-mid Array will operate with the frequency resolution of 20 kHz and 15,000 channels \citep{SKA}. Measuring the flux in bins of 100 kHz and targeting HI galaxies at $z \approx 1$ with 12hr integration per pointing and with 1 sq. degree of view,  the SKA1-mid Array is expected to complete {\em Phase 1} under a year of operation. This is presented in Fig. \ref{fig5}, which shows the time needed to complete {\em Phase 1} with 
SKA1-mid Array measuring both the redshift drift along with the flux drift. 10 years after {\em Phase 1}, the survey of the same sources needs to be repeated ({\em Phase 2}) with measurements of both redshifts and fluxes. After completion of {\em Phase 2}, the  SKA1-mid Array will provide the direct measurement of the cosmological expansion. 

\begin{figure}
\centering
\includegraphics[scale=0.4]{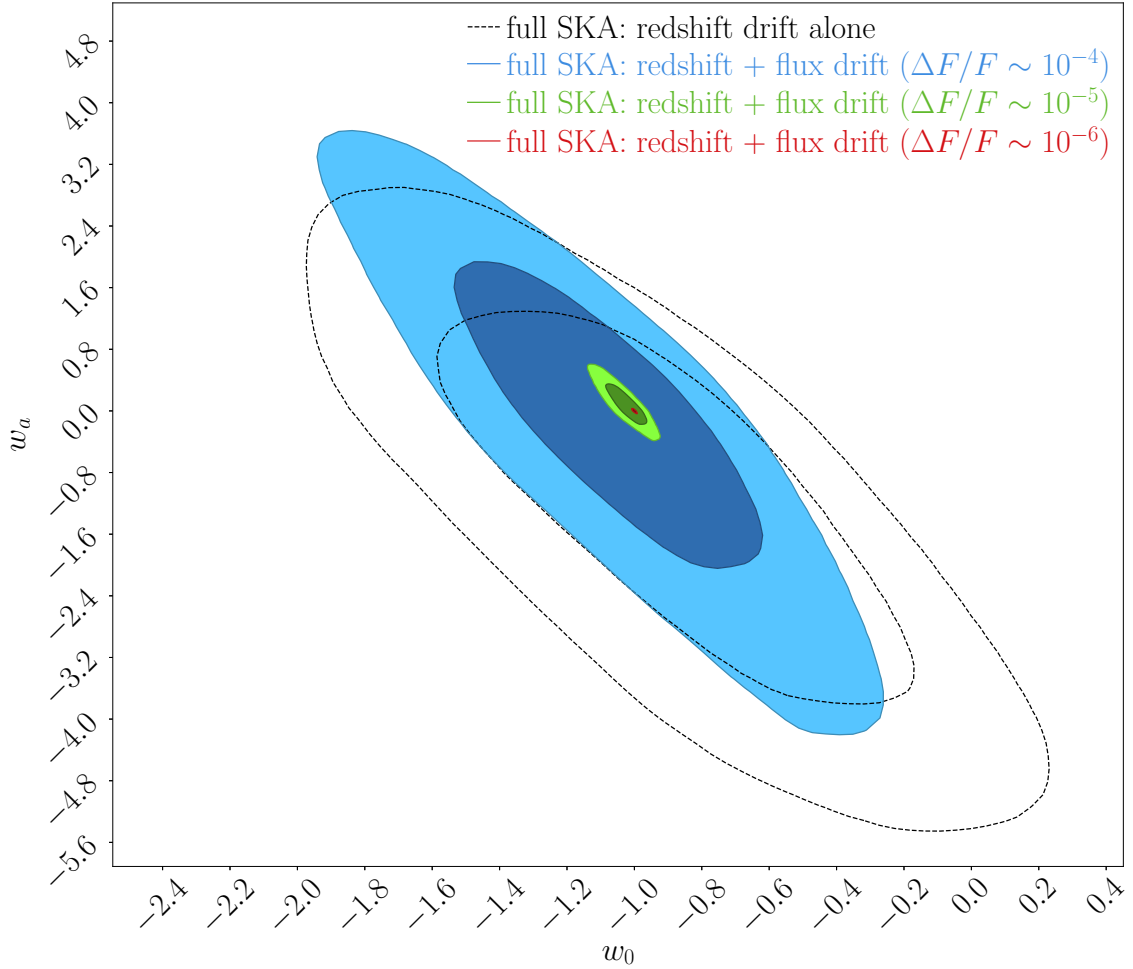}
\caption{\label{fig7}Constraints on the parameters 
 $w_0$ and $w_a$ of the equation of state of dark energy ($68\%$ and $95\%$ confidence regions) expected from the full SKA based on {\em drift effects} only. The dashed line shows constraints from the {\em redshift drift} alone. If the stability of radio galaxies is at the level of $\Delta F_\nu/F_\nu \sim 10^{-4}$, $\Delta F_\nu/F_\nu \sim 10^{-5}$, and $\Delta F_\nu/F_\nu \sim 10^{-6}$
then the expected constraints based on the {\em redshift drift} and {\em flux drift} are presented in blue, green, and red, respectively.}
\end{figure}

This however, relies on the assumption that radio galaxies are stable at the level of $\Delta F_\nu/F_\nu \sim 10^{-6}$. If the HI galaxies are not stable at this level, but have some random variability at the level of $10^{-5}-10^{-4}$, then the SKA1-mid Array will not be able to detect the expansion of the universe before the full SKA.
In the case when the radio galaxies are variable at the level of $10^{-4}$ even with the full SKA each of the observing phases will take a few hundred days to complete. 
Thus, including the flux drift will not speed up the observing phases, but it will improve the cosmological constraints, which is presented in Fig.~\ref{fig7}.

Figure~\ref{fig7} shows expected constraints on the parameters of the equation of state of dark energy. The mock data was generated based on the number of HI galaxies
expected to be observed by the SKA, as given by eq. (\ref{HIska}).
The equation of state of dark energy is parametrised in terms of the parameter $w$, which is given by the ratio of pressure to energy density
\begin{equation}
    w = \frac{p_{{\rm DE}}}{\rho_{{\rm DE}} c^2}.
\end{equation}
In the case of  the cosmological constant $w = -1$. If $w\ne-1$ then dark energy evolves, and its evolution follows from the Friedmann equations and the equation of state. In general case the equation of state can also evolve in time. In the linear series (in terms of the scale factor $a$) the equation of state can be written as

\begin{equation}
   w = w_0 + (1-a) w_a, 
\end{equation}
where $w_0$ and $w_a$ are constants, and the factor $(1-a)$ ensures that energy density of dark energy does not diverge as $a\to 0$ \cite{2003PhRvL..90i1301L}.

As seen in Fig.~\ref{fig7}, even if the flux stability is at the level of 
$\Delta F_\nu/F_\nu \sim 10^{-4}$ the constraints on the equation of state improve. In terms of the Figure of Merit (inverse of the area contained within the 95\% confidence contours of $w_0 - w_a$ \cite{2006astro.ph..9591A}) the constraints improve by a factor of 2.
In terms of the constraints on the equation of state of dark energy, the expected constraints in the year 2040 do not seem impressive if compared to the present-day constraints \cite{2016A&A...594A..13P}. However, one should keep in mind that these are the constraints from the {\em drift effects} alone, without combining them with other cosmological probes available in the year 2040 \cite{2018EPJC...78...11L}.
In addition, if the flux stability is better than $\Delta F_\nu/F_\nu \sim 10^{-4}$, then the constraints improve significantly. In terms of the Figure of Merit, these constraints increase  to  7.5  and 1,400 for $\Delta F_\nu/F_\nu \sim 10^{-5}$  and $\Delta F_\nu/F_\nu \sim 10^{-6}$  respectively. This shows that if only we can identify targets sufficiently stable 
in terms of their flux, then the measurement of the flux drift can become a useful tool in cosmological studies.

\section{Conclusions}\label{conclusions}

The analysis presented in this paper focused on `real-time' observational signatures of the cosmic expansion. Due to this expansion, the observed properties of cosmological sources change with time. These include: change of redshift, change of distance, change on flux, change of angular position, as well as, the angular size \citep{1962ApJ...136..319S,1998ApJ...499L.111L,1962ApJ...136..334M,2007MNRAS.382.1623B,2010MNRAS.402..650K,2008PhRvD..77b1301U,2010PhRvD..81d3522Q,2011PhRvD..83h3503K,2018JCAP...03..012K,2018arXiv181110284G,2015APh....62..195K}. All of these effects are of very small amplitude and will require very sensitive instruments, with observational projects lasting at least a decade.

Despite the wealth of different {\em drift effects},
in the literature only the redshift drift has been of the primary focus, and it has been shown that it is only with very large instruments such as SKA and ELT that we will finally reach sufficient sensitivity to detect the redshift drift.
Both ELT and SKA are expect to be operational around the year 2030, which means that the earliest the  redshift drift could to be detected is around the year 2040.

In this paper, apart from the {\em redshift drift} we also discussed the {\em flux drift}. The flux drift is more challenging to observe as it requires a sufficiently large sample of radio galaxies that are intrinsically stable (in terms of flux) over the period of 10 years.
We showed that when the observational project focuses on both of these  drift effects (i.e. redshift drift + flux drift)
detectibility is boosted.
 If the stability of radio sources at the level of $10^{-6}$ 
then the SKA1-mid Array should be able to detect the drift effects before the ELT and the full SKA.  This means that the direct detection of the cosmic expansion could be achieved by mid 2030.
This however requires stability of flux at the level of $\Delta F_\nu/F_\nu \sim 10^{-6}$. If the galaxies are variable at a level higher than $\Delta F_\nu/F_\nu \sim 10^{-6}$ then 
the SKA1-mid Array will not be able to detect the cosmic expansion before the full SKA. However, if only radio galaxies are stable at the level of $\Delta F_\nu/F_\nu \sim 10^{-4}$ (or better) then the inclusion of the flux drift into the analysis will significantly  improve the constraints on cosmological parameters, including the dark energy equation of state (cf. Fig. \ref{fig7}).

This paper concludes that the {\em flux drift} could become a useful cosmological tool and we advocate for utilising this tool in future cosmological studies.

\acknowledgments

KB acknowledges support from the Australian Research Council through his Future Fellowship FT140101270. CW thanks the School of Astronomy and Space Science, Nanjing University for financial support of his research at the University of Sydney as part of the Nanjing Exchange Program.
The plots were created using  \texttt{Gnuplot}\footnote{\url{http://gnuplot.info/}}, \texttt{Matplotlib}\footnote{\url{https://matplotlib.org/}}, \texttt{Healpy}\footnote{\url{https://healpy.readthedocs.io}} and \texttt{ChainConsumer}\footnote{\url{https://samreay.github.io/ChainConsumer/}  \cite{Hinton2016}}.

\bibliography{bib} 
\bibliographystyle{JHEP} 
\end{document}